\documentclass[reprint, superscriptaddress, amsmath, amssymb, aps, pra, nofootinbib, widetext]{revtex4-2}

\usepackage[utf8]{inputenc}
\usepackage{graphicx}
\usepackage{hyperref}
\usepackage{xcolor}
\usepackage{braket}
\usepackage{bm}
\usepackage[format=plain]{caption}
\usepackage{subcaption}
\begin{document}
\title{Variational quantum simulation of critical Ising model with symmetry averaging}

\author{Troy J. Sewell}
\email{tjsewell@umd.edu}
\affiliation{Joint Center for Quantum Information and Computer Science, College Park, MD, 20742}
\affiliation{University of Maryland, College Park, MD, 20742}

\author{Ning Bao}
\affiliation{Computational Science Initiative, Brookhaven National Lab, Upton, NY, 11973}

\author{Stephen P. Jordan}
\affiliation{Microsoft, Redmond, WA 98052}
\affiliation{University of Maryland, College Park, MD, 20742}

\begin{abstract}
	Here, we investigate the use of deep multi-scale entanglement renormalization (DMERA) circuits as a variational ansatz. We use the exactly-solvable one-dimensional critical transverse-field Ising model as a testbed. Numerically exact simulation of the quantum circuit ansatz can in this case be carried out to hundreds of qubits by exploiting efficient classical algorithms for simulating matchgate circuits. We find that, for this system, DMERA strongly outperforms a standard QAOA-style ansatz, and that a major source of systematic error in correlation functions approximated using DMERA is the breaking of the translational and Kramers-Wannier symmetries of the transverse-field Ising model. We are able to reduce this error by up to four orders of magnitude by symmetry averaging, without incurring additional cost in qubits or circuit depth. We propose that this technique for mitigating systematic error could be applied to NISQ simulations of physical systems with other symmetries.
	\end{abstract}
\maketitle 

\section{Introduction}

The multi-scale entanglement renormalization ansatz (MERA) tensor networks represent quantum states on a $d$ dimensional lattice using a $d+1$ dimensional representation. The extra dimension can be interpreted as scale, with slices of the network along this dimension corresponding to successively coarse grained states. In one dimension, the tree-like MERA tensor network can represent ground states of critical systems, reproducing polynomially decaying correlation functions and logarithmic scaling of subsystem entanglement entropy \cite{MERA, EV09, critical_mera}. In the case where the local tensors are fermionic Gaussian unitaries, the networks can viewed as a wavelet transform on fermion operators and can be rigorously shown to support good approximations of local free-fermion ground states \cite{ERwav, Haegeman_2018, witteveen2019quantum}.

The local unitary structure of a MERA tensor network may be interpreted as a quantum circuit which introduces further UV degrees of freedom scale-by-scale to entangle qubits of the target ground state. However, on a quantum computer it is more natural to allow increased circuit depth rather than local bond dimension as a means of increasing expressivity of an ansatz, giving rise to a class of quantum circuits known as DMERA \cite{DMERA} which could be used as a variational circuit for ground state preparation. Of particular interest for near-term application is that local observables and correlation functions of ground states prepared by DMERA circuits feature an inherent resilience to local noise stemming from circuit topology \cite{DMERA}. It is also possible to prepare subregions of DMERA states on quantum computers much smaller than the total system size, which may help to leverage the capabilities of small quantum devices \cite{preskill_quantum_2018}. This has been demonstrated on an ion-trap quantum computer, where the critical ground state subregion can be prepared as a robust fixed-point state of a local quantum channel derived from a DMERA quantum circuit \cite{ERprep}.

MERA tensor networks can be contracted in polynomial time classically. However, the polynomial scaling with bond dimension is quite severe, \emph{e.g.} $O(\chi^8)$ for 1D and $O(\chi^{16})$ for 2D using the schemes proposed in \cite{EV09}. Thus direct execution of MERA on quantum computers can yield large polynomial speedups, and can additionally serve as a method for preparing initial states in the context of a quantum algorithm for simulating quantum dynamics.

This work investigates the feasibility of multi-scale circuits for variational ground state preparation on quantum computers in the fashion of a variational quantum eigensolver (VQE) \cite{Peruzzo_2014, VQE_review}, or more generally an ansatz circuit for some variational quantum algorithm (VQA) \cite{McClean_2016}, where some family of parameterized circuits are minimized according to some Hamiltonian energy in hopes that a good approximation of the ground state can be found. {Another MERA-inspired quantum circuit ansatz for variational ground state preparation was studied in Ref. \cite{Barthel_MERA}, however, the circuit ansatz studied in the present work more directly follows the constructions of Refs. \cite{DMERA, ERwav, Haegeman_2018}. }

 { Conformal field theories and the use of renormalization are of central importance to the study of quantum field theories, which may generally be viewed as deformations from critical fixed points of a renormalization group flow. The prospect to simulate conformal field theories with critical lattice models and including the use of renormalization theory is seen as an important step in the ability to handle more general study of field theory simulations on quantum computers \cite{CFTsim}. Ground states of critical systems feature large correlation lengths which are challenging to reproduce using a more locally entangled ansatz such as matrix product states or short depth quantum circuits. Scale invariance, however, makes these states somewhat simpler to describe using a multi-scale ansatz which may incorporate this symmetry directly into the circuit parameterization. Gapped states may be better able to be represented by an ansatz with local correlations, but may also be prepared using a multi-scale circuit with parameters that differ between scales, and using fewer scales overall due to the lack of long range correlations.}
 
\begin{figure}[t]
   \begin{subfigure}{\linewidth}
   \centering
  \includegraphics[width=\linewidth]{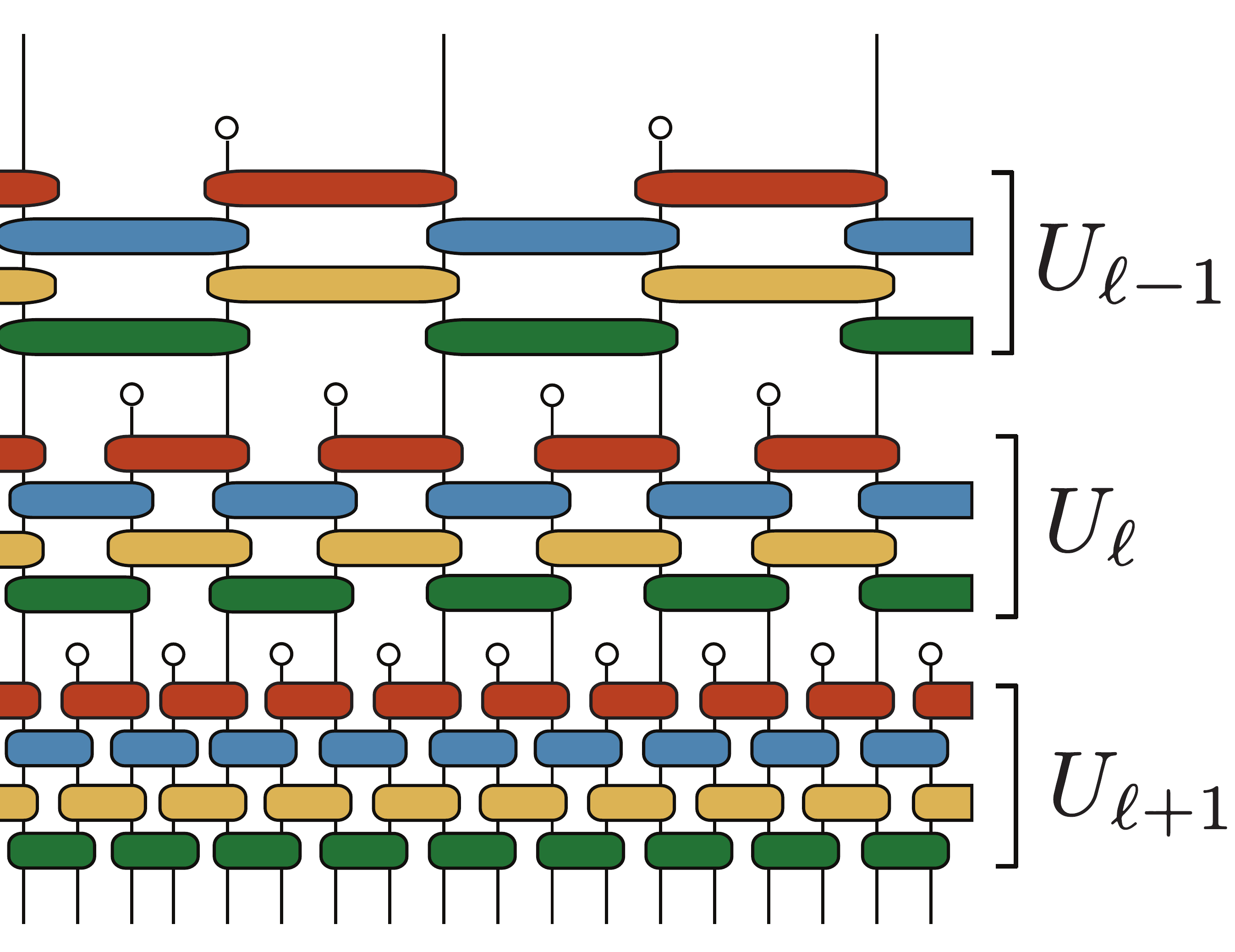}
    \caption{}
 \end{subfigure}
   \begin{subfigure}{\linewidth}
   \centering
  \includegraphics[width=\linewidth]{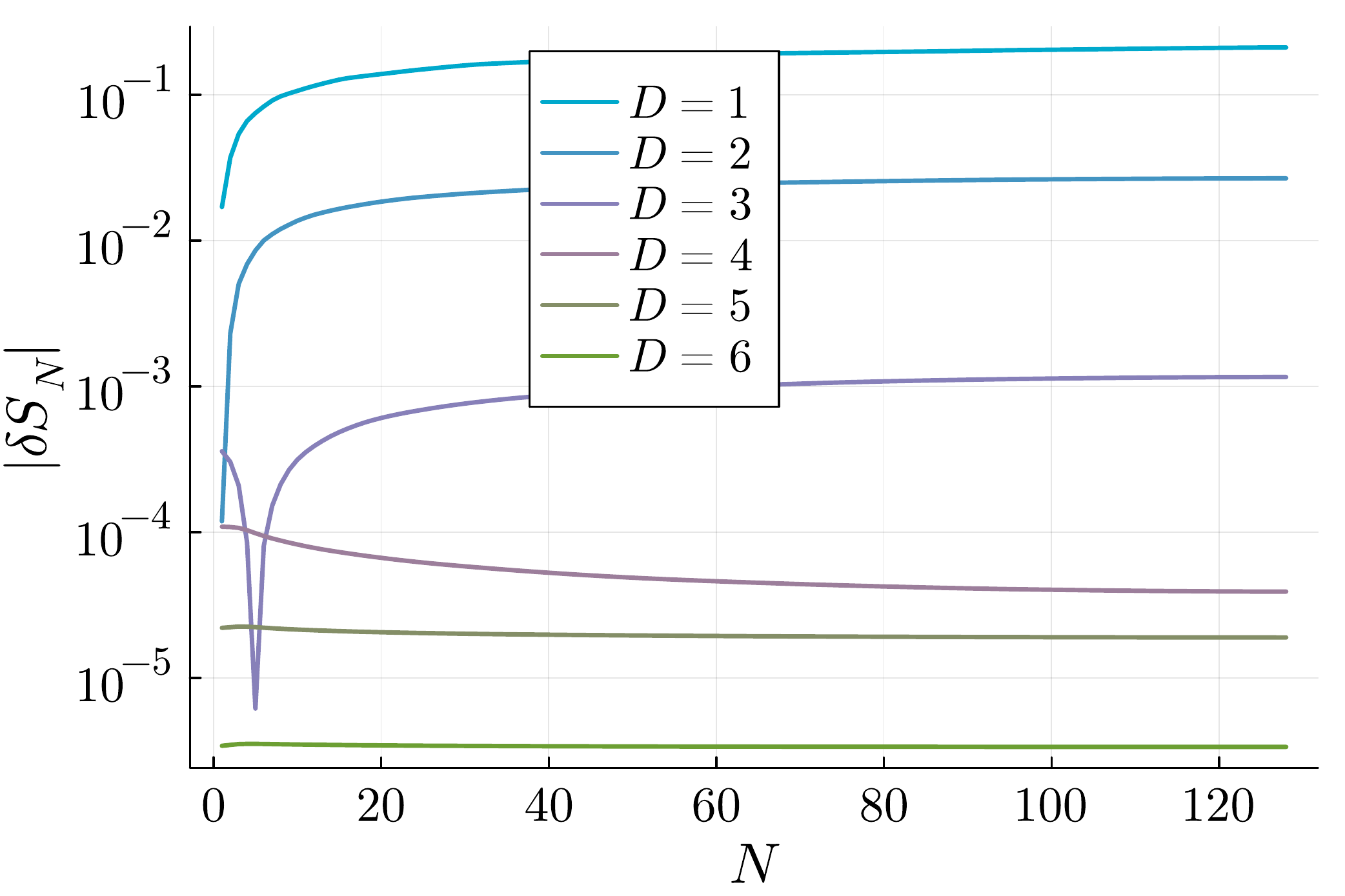}
    \caption{}
 \end{subfigure}
        \caption{\\ \textbf{(a):} State preparation by successive application of $D=4$ scale transformation circuits $U_\ell (\theta)$ with new qubits initialized in the state $\ket{0}$. Colors show the identity of local gate parameters imposed to replicate approximate translation and scaling symmetry.\\
        \textbf{(b):} Relative error in average entanglement entropy of $N$ qubit sybsystems for an $L=256$ qubit state, which is exponentially small in $D$. The subsystem entropy scales logarithmically with $N$ for critical ground states in one dimension, with DMERA circuits featuring an excess of entropy for $D<3$ and a small entropy deficit for $D>3$.  }
        \label{fig:highlights}
\end{figure}

We benchmark the viability of the DMERA ansatz for variational state preparation by numerically optimizing for approximate ground states of the critical Ising spin chain in one dimension. We are able to find high fidelity ground state approximations using relatively low circuit depth $D$ of each scaling transformation. The ansatz states and local energy density converge to their exact values exponentially with $D$, with relative error in the energy density below $10^{-8}$ for $D=6$. Key features of critical ground states in one dimension such as polynomially decaying correlation functions and logarithmic scaling subsystem entanglement entropy are found on average. See Fig. \ref{fig:highlights} for a representation of the multi-scale variational circuit with $D=4$.

{ Symmetry-averaging is also used to improve the systematic error of local observable expectation values. This can be implemented in hybrid variational schemes by using classical post processing to average observables which are related by symmetries which are explicitly broken in the circuit ansatz. In the case of the critical Ising model spatial translation and Kramiers-Wannier symmetry are used, although we also note that these symmetries are still approximately reproduced in the ansatz states themselves, with smaller variance over these symmetry groups for increasing $D$. Similar ideas have been studied in the context of Refs. \cite{Seki_2020, symm_2}, where symmetry projection is used in post-processing to improve broken symmetries of ansatz states.}

\begin{figure}[t]

   \begin{subfigure}{\linewidth}
  \includegraphics[width=\linewidth]{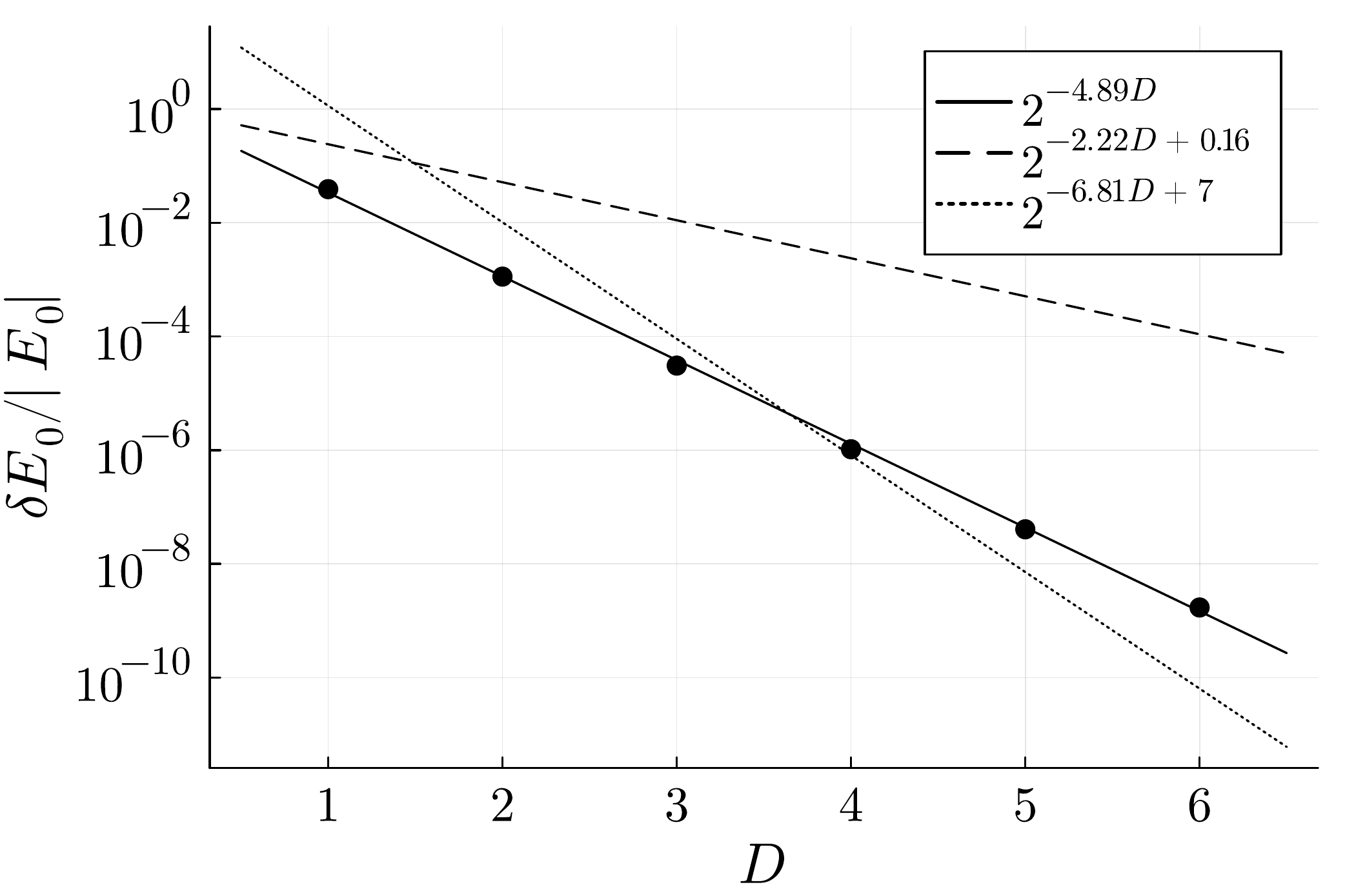}
  \caption{}
 \end{subfigure}\hfill
   \begin{subfigure}{\linewidth}
  \includegraphics[width=\linewidth]{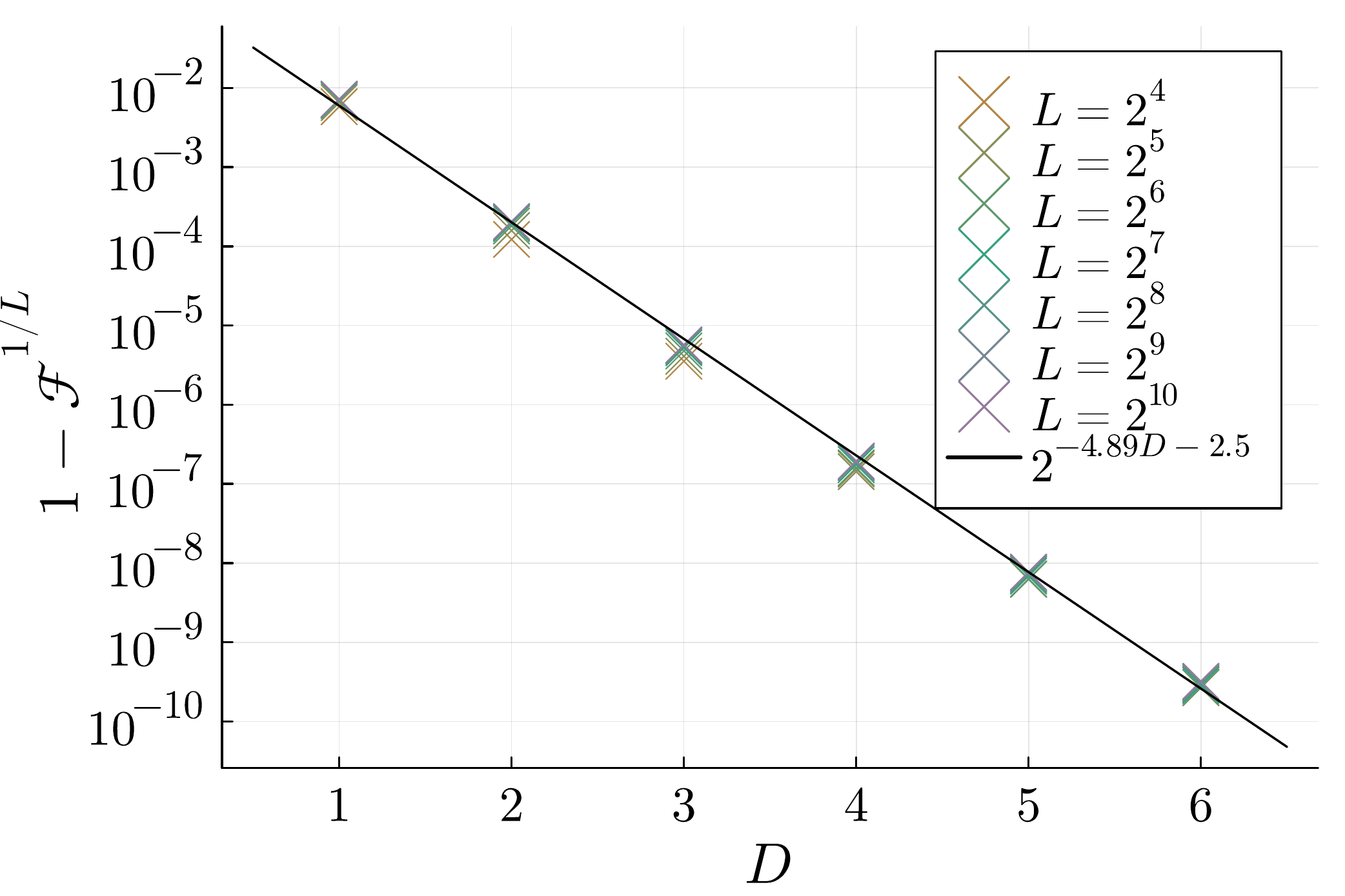}
  \caption{}
 \end{subfigure}
  \caption{\\ \textbf{(a):} Relative error of variational fixed-point ground state energy for channels of depth $D$, using the infinite volume energy density of $-4/\pi$. Trend lines showing the exponential scaling of energy density error for our ansatz (solid) as well as those from the analytic wavelet construction of \cite{Haegeman_2018}(dashed) and the numerically optimized non-Gaussian MERA in \cite{critical_mera} (dotted). \\ \textbf{(b):} Normalized state infidelity $1 - \mathcal{F}^{1/L}$ of states prepared by depth $D$ scaling transformation for ground states of various system sizes $L$. \\ Both quantities appear to decay exponentially with $D$ with a coefficient of approximately $-4.89$.}
\label{fig: energy-fidelity}
\end{figure}

\section{Circuit Construction} \label{s:circuit-ansatz}

We consider states which double the number of qubits with each scaling transformation, interleaving new qubits in the zero state between qubits from each of the previous layers. After $\ell$ scaling circuits we have a state on $L = 2^\ell$ qubits. Our goal is to prepare an approximation to the ground state of a corresponding critical Ising Hamiltonian. Each transformation is a quantum circuit $U(\theta)_\ell$ of depth $D$, parameterized in terms of angles $\theta$.
\begin{equation}
  \label{eq:dmera-scaling}
  \ket{\psi_{\ell+1}} = U_{\ell}(\theta)\Big(\ket{ \psi_\ell} \otimes \ket 0^{\otimes 2^\ell}\Big)\;.
\end{equation}
Approximate translation and scale invariance of the state are imposed by having each scale transformation be a periodic brickwork circuit invariant under translation by two sites, and constraining each scaling transformation circuit at different layers to be made up of the same gates, albeit acting on different numbers of qubits. The approximate translation symmetry reduces the total number of variational parameters from scaling with the total number of gates, $O(D L)$, to only with total circuit depth $O(D \log L)$. The approximate scale symmetry further reduces the number of circuit parameters down to only $O(D)$, the depth of a single scaling transformation circuit. (See fig \ref{fig:highlights}).

These symmetries cannot be imposed exactly due to the discrete nature of the gates and scaling transformations, so the exact symmetries are broken by the ansatz state, yet retained as approximate symmetries. Deviations in local observables are exponentially small in $D$ but constitute a major source of error for experimentally realistic values of $D$, \emph{e.g.} $D=1,2,\ldots,6$ as studied here. However, as shown in Fig. ~\ref{fig:2pt}, the error in two-point correlation functions {related by Kramiers-Wannier symmetry are nearly out of phase with each other.} Consequently, this source of error can be reduced by averaging expectation values over this known symmetries. This symmetry averaging lowers error in the correlation functions by approximately two orders of magnitude, yielding relative error below $10^{-7}$ for correlation functions using the $D=6$ state. This improvement is consistent across both of the example Hamiltonians considered in this work (eq. \ref{eq:ham1} and \ref{eq:ham3}), as shown in Figs. ~\ref{fig:2pt} and \ref{fig:sym-ratio}.

In \cite{ERwav, Haegeman_2018, witteveen2019quantum}, wavelet transformations on fermionic modes are used to construct approximations to free fermion ground states. These constructions use knowledge of the underlying physics to analytically construct the desired wavelet and achieve accuracy exponential in the order of the wavelet (corresponding to circuit depth $D$). In this work we instead determine the variational parameters in our DMERA ansatz by numerical optimization. Inspired by the free-fermion MERA construcions of \cite{ERwav} we take
\begin{equation}
\label{eq:local-gate}
u(x,y) = \begin{bmatrix}
\cos(x) & 0 & 0 & \sin(x) \\
 0& \cos(y) & \sin(y) & 0 \\
 0& -\sin(y) & \cos(y) & 0 \\
 -\sin(x)& 0 & 0 & \cos(x) \\
\end{bmatrix}
\end{equation}
as the local gates in our circuit, where the two variational parameters specify the rotation on the odd and even parity subspaces of the pair of qubits. Because we use real-valued parity-conserving gates, the anstatz states will always be parity-even states with time-reflection symmetry. 

Similarly to \cite{Haegeman_2018}, we find that variationally optimized parameters achieve substantially more accurate energy densities than analytically constructed parameters, even though the latter become exact in the limit of DMERA of infinite depth or MERA of infinite bond dimension.

\section{Model and Variational Optimization}\label{s:var}

We focus on the transverse-field Ising model at criticality, namely

\begin{equation}
	\label{eq:ham1}
	H_I = -\sum_{j=1}^L X_j X_{j+1} + Z_j.
\end{equation}

This spin chain is known to be well described by a conformal field theory with $c=1/2$ at low energies, but is also integrable due to the Jordan Wigner duality relating it to the following free fermion model

\begin{align}
  \label{eq:ham2}
  H = i \sum_{j=1}^{2L} \gamma_j \gamma_{j+1},
\end{align}
where $\gamma_j$ are Majorana operators. { In this convention there are $2L$ Majorana operators for $L$ spatial sites, with operators $\gamma_{2j}$ and $\gamma_{2j-1}$ together comprise the fermion at spatial site $j$. At the critical point the coupling between Majorana operators at the same spatial site and neighboring sites are equal, which is sometimes referred to has the "half-shift symmetry", and allows for the simple description of the Hamiltonian in Eqn. \eqref{eq:ham2}.} For the spin Hamiltonian on an even number of sites with periodic boundary conditions, the ground state has even parity and is equivalent to the ground state of the Majorana Hamiltonian with anti-periodic boundary conditions, so $\gamma_{2L+1} = - \gamma_1$.

To achieve efficient numerical simulations we exploit two properties. First, the gates of eq. \ref{eq:local-gate} are matchgates, which allows us to simulate their action on Gaussian fermion states efficiently as linear transformations on Majorana operators \cite{Jozsa_2008, FLO}. By constraining to matchgate tensors, tensor networks may be used to represent Gaussian fermion operations and states \cite{Schuch_2019, Jahn_2017}. We use these techniques to compute two-point correlation functions out to large distances such as 512 lattice spacings. Secondly, the expectation values of local operators in MERA and DMERA states can be accelerated by exploiting the constant-width causal cones imposed by these circuits. Local reduced density matrices of the global pure state defined by the ansatz can be obtained as output from iterated quantum channels on a constant number of qubits. These local density matrices are all that is needed to compute the energy of the variational state under a local Hamiltonian. As the number of layers in the MERA ansatz is increased the density matrix converges exponentially to the fixed point of this channel, which can be easily computed, as described in \cite{ERprep}. We tune variational parameters by minimizing the energy density of this fixed-point state.

For sufficiently shallow circuits, we find well-optimized variational parameters using gradient descent and L-BFGS with restarts. For higher depth circuits we obtain variational parameters by starting from an optimized circuit for a given value of $D$ with an additional row of gates inserted at the end of the circuit with parameters initialized near the identity, and then re-optimizing to obtain a solution for $D+1$. Similarly, we may alternatively insert two rows of near-identity gates anywhere within the scaling circuit to obtain a solution for $D+2$. {For the larger depth circuits, the optimization by adding additional parameters to a smaller solution was better at finding good minima. Random initialization of all parameters was less likely to find an energy minimum any lower than found previously with shorter depth circuits. We therefore relied on this technique of bootstrapping a better $D+1$ minimum from a depth $D$ minimum in order to find circuits which continued to show exponentially small error in the energy with circuit depth.

By using the total energy to estimate the average energy density, we are implicitly averaging over symmetries of the Hamiltonian, since the full symmetry group we consider for the critical Ising model relates each local term in the Hamiltonian to every other. By not preparing the whole state and only measuring some terms of the Hamiltonian, we run the risk of minimizing the energy in a non-variational way, i.e. minimizing an estimate of the global energy below what is possible for the ground state, while inadvertently raising the energy for terms not measured because the global energy still cannot be below that of the ground state. While this could happen in principle, the spatial distribution of expectation values does not have a large variance due to the approximate translation symmetry still found in the circuit ansatz, and so this likely would not happen in a catastrophic way in practice. }

Imposing swap-symmetry on the local gates would constrain the odd-parity rotation angle $y=0$. For the isometry gates only the output legs are swapped, so instead this angle would bet set to $y =- \pi / 4$, resulting in an ansatz state which is symmetric with respect to spatial reflections. This choice of parameters is optimal for the $D=1$ case. However, for $D>1$ we find that by relaxing this constraint, reflection-symmetry broken states can be prepared using the same circuit depth which have lower energy and higher fidelity. By transforming the parameters $y \rightarrow -y$ for non-isometry gates and $y \rightarrow \pi /2 -y$, a reflected ansatz state with the same energy density and fidelity is prepared.

The circuit parameters we use throughout this paper for $D \leq 6$ can be found in the Appendix. The corresponding error in energy density appears to be exponentially small in $D$ with a scaling coefficient of $-4.89$, seen in Fig. \ref{fig: energy-fidelity}. We compare this fit line to two similar multi-scale ground state representations for the Ising model, the wavelet based construction found in Ref. \cite{Haegeman_2018} which uses a very similar circuit ansatz but constructs the ideal parameters analytically rather than through energy minimization, and the more traditional MERA ternsor network states in Ref. \cite{EV09} found through numerical optimization of the energy. In order to compare these different methods, we note that local gates of the circuit based methods can be grouped together to form a MERA tensor network of bond dimension $\chi = 2^{D-1}$. So, while the MERA from \cite{EV09} does not have a circuit depth $D$, we will use this equivalence in order to compare ansatz states of similar complexity. That being said, the circuit based constructions of this work and \cite{Haegeman_2018} are much more constrained ansatz states compared to a generic MERA of bond dimension $\chi = 2^{D-1}$, due to the local circuit structure. 

We find that while maintaining much of the simplicity of the wavelet construction of \cite{Haegeman_2018}, our numerically optimized states are able to outperform those constructions at equal circuit depth and appear to scale with a larger scaling coefficient. Our ansatz states even outperform the general MERA states of \cite{EV09} for short depth circuits, but are beaten at larger bond dimensions, which should be expected due circuits being much more constrained than a MERA state of equivalent bond dimension. 

\section{Numerical Benchmarking} \label{s:benchmarking}

\begin{figure}
\centering
   \begin{subfigure}{\linewidth}
  \includegraphics[width=\linewidth]{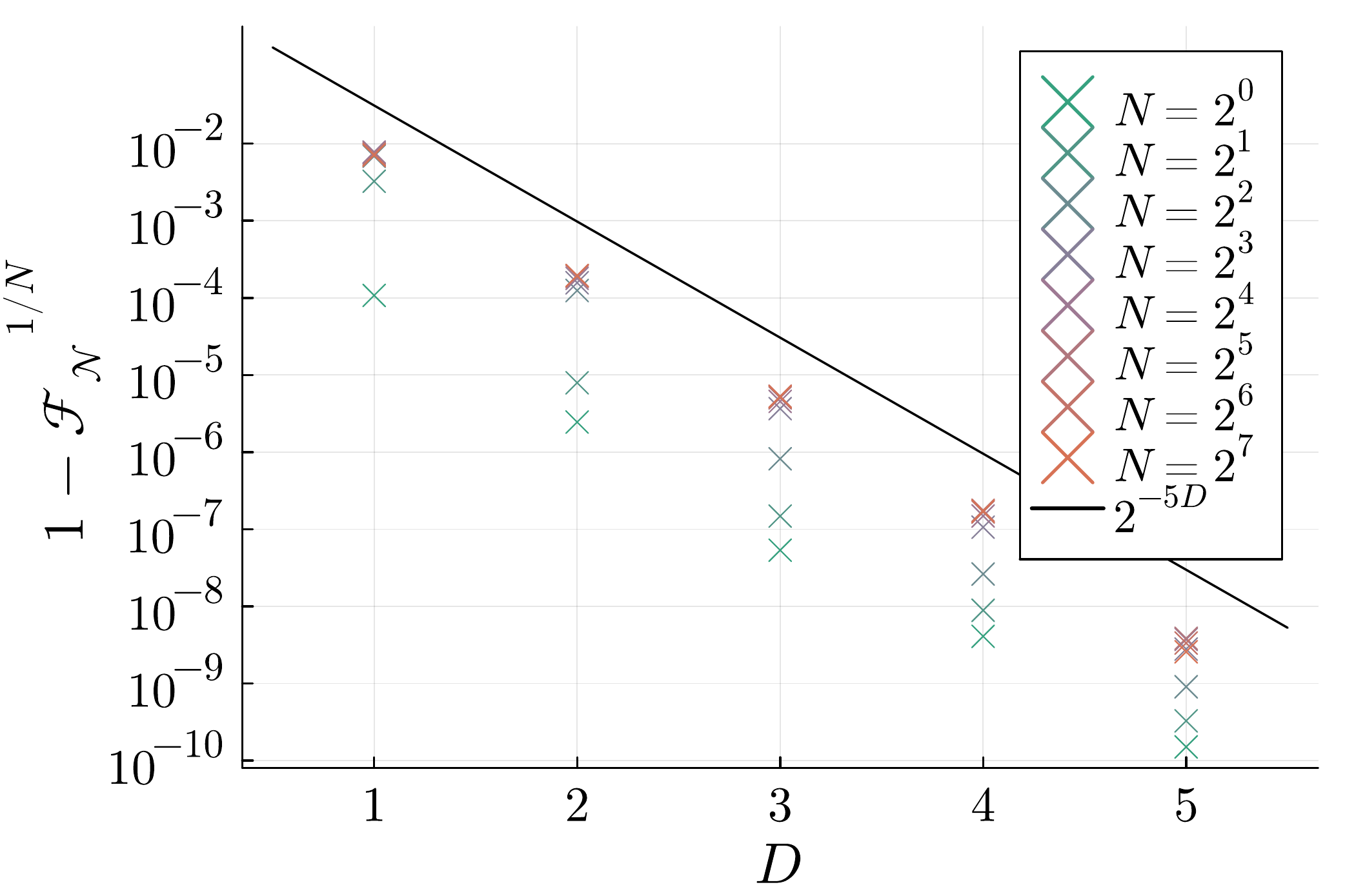}
 \end{subfigure}
  \caption{Normalized infidelity of $N$ qubit subsystems of a $L=256$ qubit state, averaged over different subsystem locations.}
  \label{fig:subinfid}
\end{figure}

Having minimized the energy density for circuits with $D \leq 6$, we can compare our ansatz states with the true ground state for a variety of other measures. We compare the state fidelity and subsystem fidelities for different (sub)system sizes and circuit depths as well as logarithmic scaling of subsystem entropy. Finally we look at the two-point correlation functions and show how averaging local observables over symmetry groups broken by the circuit ansatz can greatly improve the estimation accuracy of these observables. 

We prepare finite sized ground states using the fixed-point optimized circuit parameters and applying some number of scaling transformation layers $\mathcal{L}$ to prepare the state on $L = 2^\mathcal{L}$ qubits. Interestingly, the infidelity of our ansatz states seem to be exponentially small in $D$, with the same scaling coefficient of $-4.89$ as the energy density, see Fig. \ref{fig: energy-fidelity}. We compare different sized systems by normalizing the state infidelity as $1- \mathcal{F}^{1/L}$, which we find clusters together ansatz states of the same $D$ but different system sizes $L$. 

In Fig. \ref{fig:subinfid} we compare the infidelity of $N$ qubit subsystems of a $L=256$ qubit system, normalized by subsystem size. This behaves similarly to that of global state fidelity with identical exponential scaling in $D$ and clustering of subsystem infidelity from the normalization. The smallest subsystems, however, have an even smaller normalized infidelity than that of the global state, and so local observables should be even more accurate than global state properties.

The relative error in average subsystem entropy is shown in Fig. \ref{fig:highlights}, showing exponential accuracy with $D$. Critical ground states in one dimension feature more long-range entanglement than area-law states, which would have constant entropy in one dimension. Instead, subsystem entropy scales logarithmically with the subsystem size which becomes a major challenge of strictly local circuits which can only entangle qubits within a finite range. Multi-scale circuits introduce entanglement at all scales and can therefore can prepare states with this scaling of subsystem entropy. In fact, the ansatz states for $D \leq 3$ have on average an excess of entropy rather than a deficit. 

\section{Averaging Broken Symmetries}

\begin{figure}
 \centering
   \begin{subfigure}{\linewidth}
  \includegraphics[width=\linewidth]{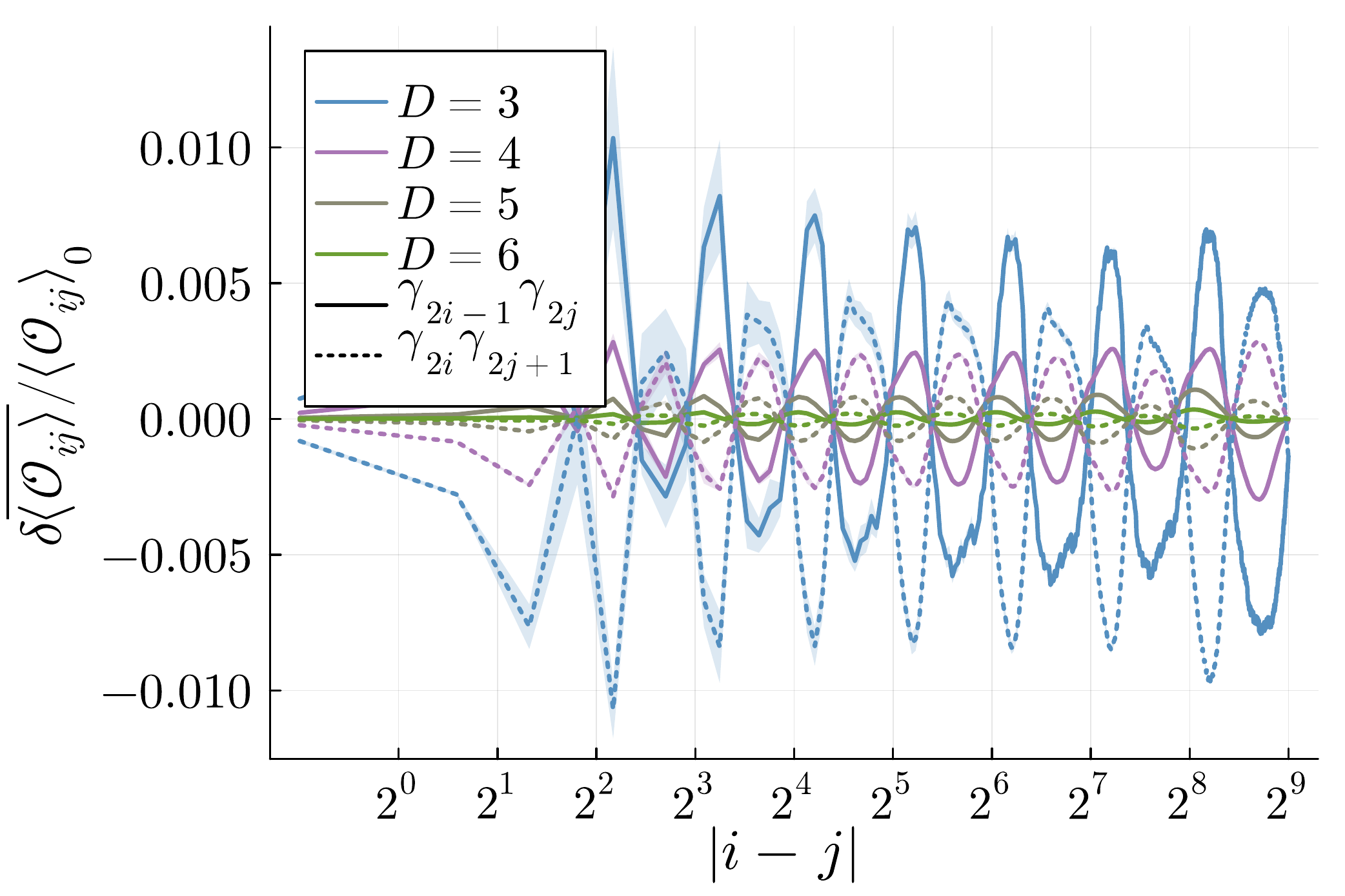}
  \caption{}
 \end{subfigure}\hfill
   \begin{subfigure}{\linewidth}
  \includegraphics[width=\linewidth]{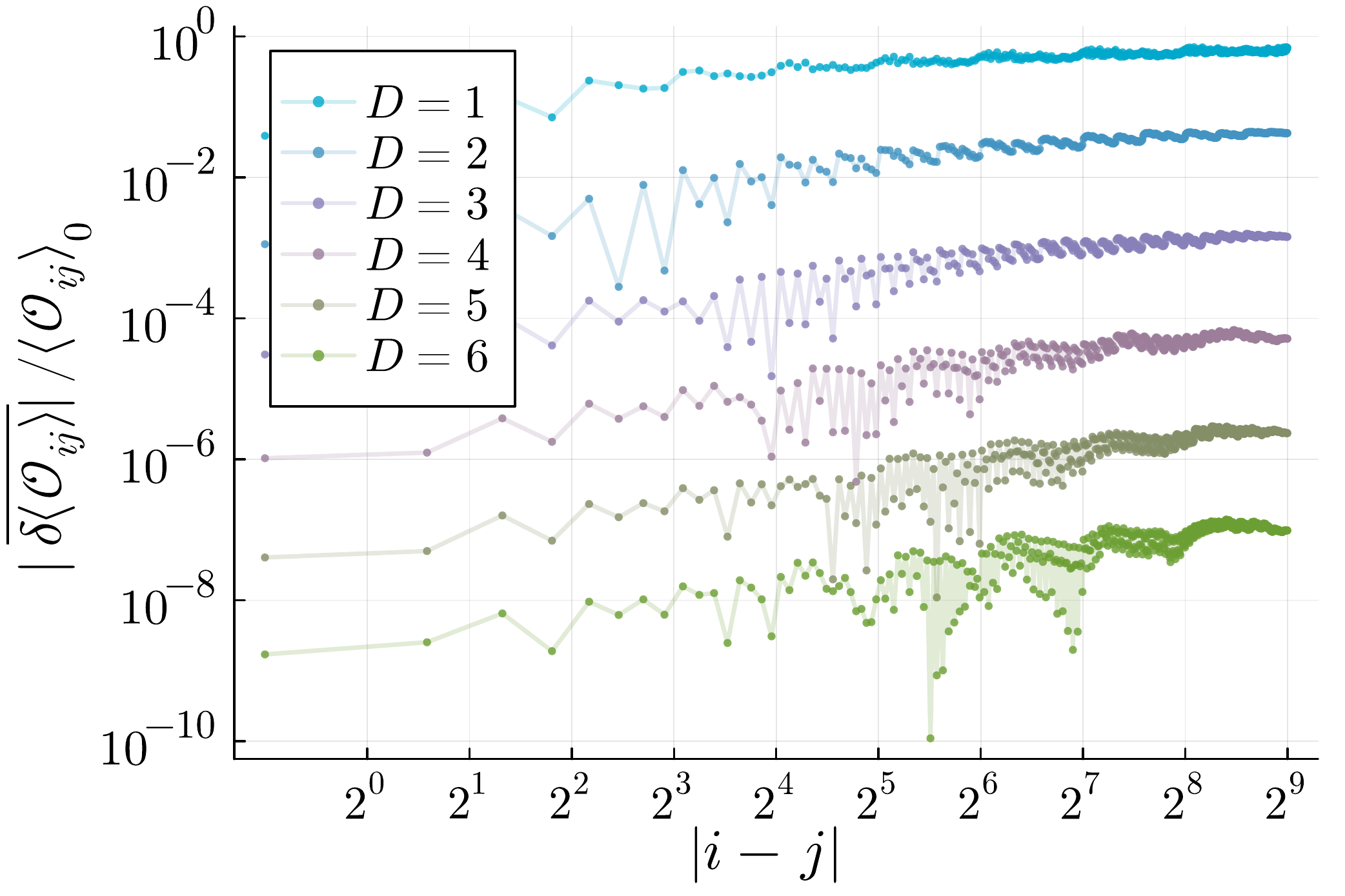}
  \caption{}
 \end{subfigure}
  \caption{\\ \textbf{(a):} Relative error in the translation-averaged expectation value of the quadratic Majorana operators $\gamma_{2i -1} \gamma_{2j}$(solid) and $\gamma_{2i} \gamma_{2j+1}$(dashed) at fixed distance $|i-j|$. These two cases are related by a Kramers-Wannier transformation, and have systematic errors which appear oscillatory on a logarithmic scale and nearly out of phase with each other.\\ \textbf{(b):} Further averaging over the Kramers-Wannier symmetry cancels much of the error of quadratic Majorana observables, reducing the error in the expecation value estimate by orders of magnitude. }
  \label{fig:2pt}
\end{figure}

Although we impose a quasi-translation invariance in the local gates of our ansatz, the actual circuit breaks this translation invariance in within circuit layer, preparing a state which is only exactly invariant under translations by $L/2$ sites. Despite this, an approximate symmetry is still maintained, so local observables related by translations have a small variance which decreases exponentially with $D$. 

In the Majorana representation there are two operators $\gamma_{2i-1}$ and $\gamma_{2i}$ at site $i$, so a translation by $n$ sites amounts to shifting the index of all Majorana operators by $2n$, i.e. $\gamma_j \rightarrow \gamma_{j + 2n}$. 

At criticality, the Ising model possesses an additional symmetry which constrains the expectation value of local $Z$ and $XX$ operators to be identical \cite{KWdual}. The Kramers-Wannier transformation maps between these sets of operators as 
\begin{equation}
Z_i \rightarrow X_i X_{i+1}, \hspace{5mm}
X_i X_{i+1} \rightarrow Z_{i+1}
\end{equation}
which is a symmetry of the Hamiltonian. This symmetry becomes more apparent in the Majorana representation, since the Jordan-Wigner transformation maps $Z_i$ to $\gamma_{2i-1} \gamma_{2i}$ and $X_i X_{i+1}$ to  $\gamma_{2i} \gamma_{2i+1}$, so the Kramers-Wannier transformation amounts to shifting the Majorana index by one and is sometimes known as a ``half-shift'' symmetry of the Majorana operators since two of these transformations amount to a spatial translation by one site. This transformation is clearly a symmetry of the Hamiltonian in Eq.~\eqref{eq:ham2}, but is broken away from criticality when the $Z$ and $XX$ operators have different coefficients.

At the top of Fig. \ref{fig:2pt} we look at the relative error in the expectation values of the two sets of quadratic Majorana operators $\gamma_{2i-1} \gamma_{2j}$ and  $\gamma_{2i} \gamma_{2j+1}$. Translation symmetry dictates that the ground state expectation values are the same for equal $i-j$, so we can average over the group of symmetries, the cyclic group on $L$ elements{, and we plot the relative error after translation-averaging in the top of Fig. \ref{fig:2pt}}. These two sets of observables decay inverse polynomially in the distance between the two Majorana operators as $|i-j|^{-1}$, and are related to each other by the Kramers-Wannier transformation. 

Interestingly, the relative error in these two sets of correlation functions appear oscillatory when plotted on a logarithmic scale in the distance $|i-j|$, which we attribute to the breaking of continuous scaling symmetry down to an approximate symmetry due to the discrete scaling transformations. The amplitude of these deviations decays with $D$, but also appears to be nearly out of phase with each other for the two sets of operators related by the Kramers-Wannier transformation. Averaging over this additional $\mathbb{Z}_2$ symmetry we can cancel much of this remaining systematic error { due to the respective deviations being nearly out of phase, resulting in a much more accurate estimation of the observables after averaging of the Kramers-Wannier symmetry, seen at the bottom of Fig. \ref{fig:2pt}.}

\begin{figure}
 \centering
   \begin{subfigure}{\linewidth}
  \includegraphics[width=\linewidth]{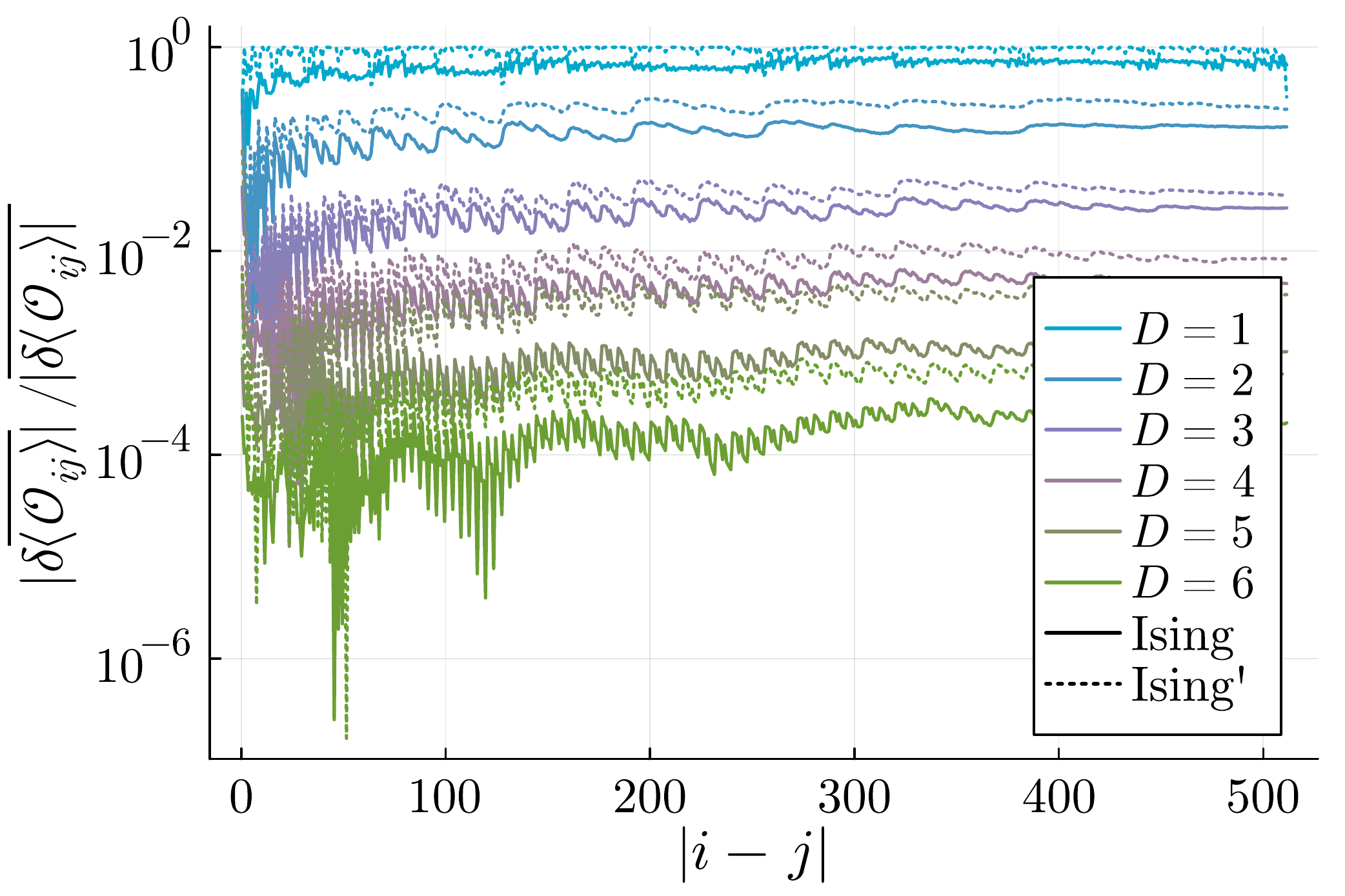}
 \end{subfigure}
  \caption{Improvement in accuracy from symmetry averaging also increases with $D$. The  ratio is plotted between the error magnitude in the symmetry-averaged correlation function, denoted by  $|\delta \overline{\langle \mathcal{O}_{ij} \rangle}|$, and the average error in particular expectation values with fixed $|i-j|$, denoted by $ \overline{|\delta\langle \mathcal{O}_{ij} \rangle|}$. Also included are the ratios for approximate ground states of the modified Ising model $H_{I'}$, which show similar improvement from symmetry averaging.  }
  \label{fig:sym-ratio}
\end{figure}

We can assess how well this cancellation by Kramers-Wannier symmetry works by comparing the average amount of error of correlators at a fixed distance, denoted $\overline{|\delta \langle \mathcal{O}_{ij} \rangle|}$, with the amount of error in the symmetry-averaged expectation value, denoted $|\delta \overline{\langle \mathcal{O}_{ij} \rangle}|$. In Fig. \ref{fig:sym-ratio} we see that this cancellation works remarkably well, with the symmetry averaged mean showing multiple orders of magnitude better accuracy than the typical local observable. This shows that states of larger circuit depth feature both a lower amount of variance in their oscillatory expectation values as well as a higher degree of error cancellation, which compound to produce highly accurate estimators of local observables after symmetry averaging. 

In Fig. \ref{fig:sym-ratio} we also include values from circuits approximating the ground state of an alternate Ising model 
\begin{equation}
	\label{eq:ham3}
  H_{I'} = \sum_{i=1}^L -X_i X_{i+1} + X_{i-1} Z_i X_{i+1}.
\end{equation}
This Hamiltonian is related to the Ising model by a constant depth unitary {$\prod_j \frac{1}{\sqrt{2}}(-I + i X_j X_{j+1})$}, which smears the single Pauli $Z$ term into a 3-body operator, but is still equivalent to a quadratic fermion model by Jordan-Wigner duality. This other set of parameters for the $H_{I'}$ ground state feature similar degrees of error cancellation.

{While these two Hamiltonians share much of the same physics due to their equivalence up to a short-depth local transformation, the variational circuits we find to approximate their respective ground states are not made equivalent by any short depth local transformation. Both sets of circuits preparing ground state approximations to $H_I$ or $H_{I'}$ are scale-invariant variational circuits with different sets of parameters. It would be possible to prepare approximate ground states for $H_I$ by using the circuits fro $H_{I'}$ and then applying the short-depth transformation that relates the two Hamiltonians, or vice versa, but the resulting circuits are not scale-invariant due to the final circuit layer differing from all the previous ones. 

This breaking of scale symmetry is explicitly used in some applications of MERA, where an initial and final layer of tensors which differ from the scale-invariant layers are used and can be referred to as "transition layers". These may be beneficial to include explicitly because the finite volume and lattice spacing of the model already break scale-invariance at the largest and smallest scales, and so a circuit with complete scale symmetry may not be the most ideal representation of the ground state. We restrict our attention to circuits with identical variational parameters for all scaling transformations within the quantum circuit in this work, and to this extent we consider the circuits found for ground states of $H_I$ and $H_{I'}$ to give independent confirmation to the efficacy of these circuits.}

Subsystem entropy also varies with location due to lack of exact translation invariance. However, entropy is a much more coarse grained property of a state and did not display a great degree of error canceling from translation averaging. That is, the subsystem entropies were typically biased in a particular direction for all subsystems, so the average entropy was not much more accurate than that of a typical subsystem.

Here, we have specifically considered the translational invariance and Kramers-Wannier symmetries of two-point functions on a one-dimensional $N$-site translationally-invariant lattice with periodic boundary conditions. This symmetry averaging procedure can be directly generalized to arbitrary groups of symmetries, \emph{e.g.} $S_n$ symmetries on permutation-symmetric systems, the translation and rotation symmetries in a fixed number of spatial dimensions, or internal global or gauge symmetries, which could be either discrete or continuous.

Averaging over translations by a symmetry group $G$ when measuring an observable comes at no cost in qubit count or circuit depth. Rather, the requirement is for repeated executions of the circuit and measurement. Such repeated measurements are already necessary, as $1/\epsilon^2$ measurements are needed to measure an observable with statistical error of order $\epsilon$. Instead of repeating $1/\epsilon^2$ measurements on an individual choice of $G$-translation of a given observable, one can instead choose a uniformly random $G$-translation of the observable for each measurement. In the absence of any \emph{a priori} information to suggest some particular $G$-translation as preferable, this $G$-averaging will in the worst case yield equivalent accuracy to a randomly chosen but fixed $G$-translation, and in the best case can result in greatly improved accuracy due to cancellation of errors of opposite sign. The latter situation is observed in the two examples $H_I$ and $H_{I'}$ considered here. It remains for future work to investigate which models exhibit such strong benefit from cancellation and which do not.

{The only cost for symmetry averaging is potentially having to sample multiple circuits in order to prepare different subregions for measuring the collection of observables to be averaged. This makes it a viable tool for implementation on NISQ hardware, and less resource intensive than doing a full projection onto the trivial representation of the symmetry group as in Ref \cite{Seki_2020}. Since Pauli strings only have eigenvalues $\pm 1$, their variance is $1-\langle \mathcal{O} \rangle^2$ for an observable $\mathcal{O}$. The symmetry averaged expectation value will still only have measurement outcomes $\pm 1$, so aggregating the measurements from different observables will result in sample with variance $1-\langle \overline{\mathcal{O}} \rangle^2$. }

This technique is complementary to the widely-used Richardson extrapolation and randomized resampling methods introduced in \cite{errorMit}, as it mitigates systematic error induced by the ansatz rather than random error introduced by the hardware. These observations also suggest that imposing strict symmetries could be overly restrictive for a circuit ansatz. Instead, it could be more advantageous to consider ansatz states which can break these symmetries while retaining accurate symmetry-averaged expectation values. With more quantum resources available, one could consider a coherent analogue of this error mitigation scheme in which states with approximate symmetries are projected onto the proper symmetric subspace after preparation to achieve higher fidelity states. Specifically, one can project onto the $G$-invariant subspace preparing a control register in the uniform superposition, executing a controlled $G$ translation on the target state, and then measuring the control register in the Hadamard basis.

\section{Comparison to Quantum Alternating Operator Ansatz}
\label{s:local}

Another natural approach to variational state preparation is using a strictly local circuit by evolving commuting terms in the Hamiltonian in sequence to form a quantum alternating operator ansatz \cite{Hadfield_2019,farhi2014quantum}. Using a short depth ansatz makes it feasible for NISQ devices, and these circuits may be constrained to conserve symmetries of the Hamiltonian, such as preparing only translation invariant states. 
In this case, they do not benefit from symmetry averaging, unlike DMERA. This approach was used to find exact variational circuits for ground states of the critical Ising model on $2p$ spins using $p$ rounds of a QAOA ansatz \cite{Ho_2019}, and has been implemented experimentally on an ion-trap device \cite{zhu2020generation}.

Exact variational circuits are not possible with fewer than $L/2$ rounds, as this is the minimum circuit depth needed for the ansatz to ``see the whole graph'' \cite{see_graph}. That is, the past causal cone of some subregion of a local circuit ansatz contains the entire initial state only when the circuit depth is $L$ or greater, making global entanglement possible. Before a local circuit can see the whole system it also is unaware of system size. So, even though the ground state of a finite size Ising chain with periodic boundary conditions has energy density lower than the infinite volume value of $-4/\pi$, a local circuit with depth less than $L$ could not prepare a state with energy below this threshold. This also means the energy density for a local circuit ansatz will be identical for all system sizes greater than the circuit depth. 

\begin{figure}
  \centering
   \begin{subfigure}{\linewidth}
  \includegraphics[width=\linewidth]{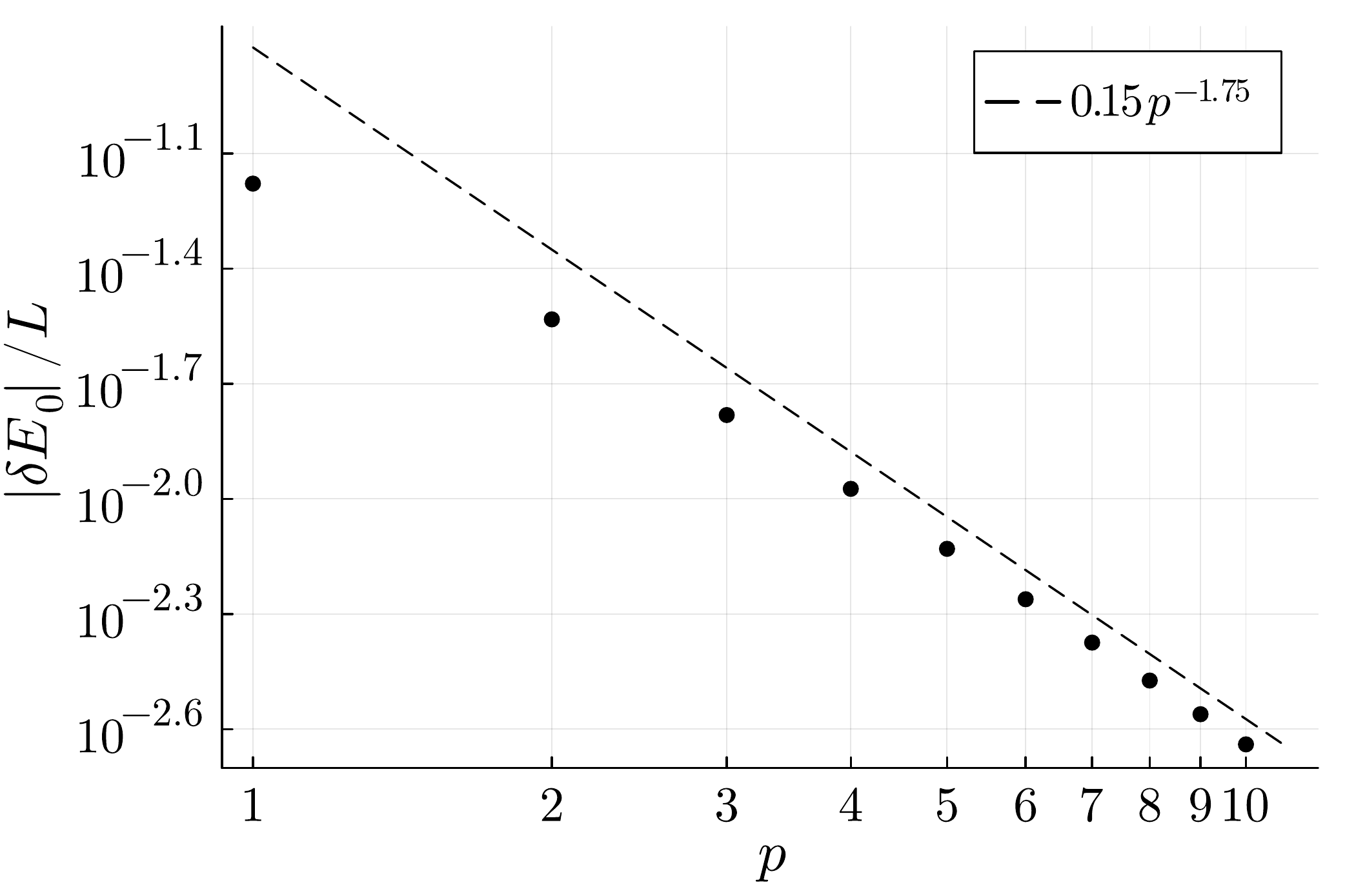}
  \caption{}
 \end{subfigure}\hfill
   \begin{subfigure}{\linewidth}
  \includegraphics[width=\linewidth]{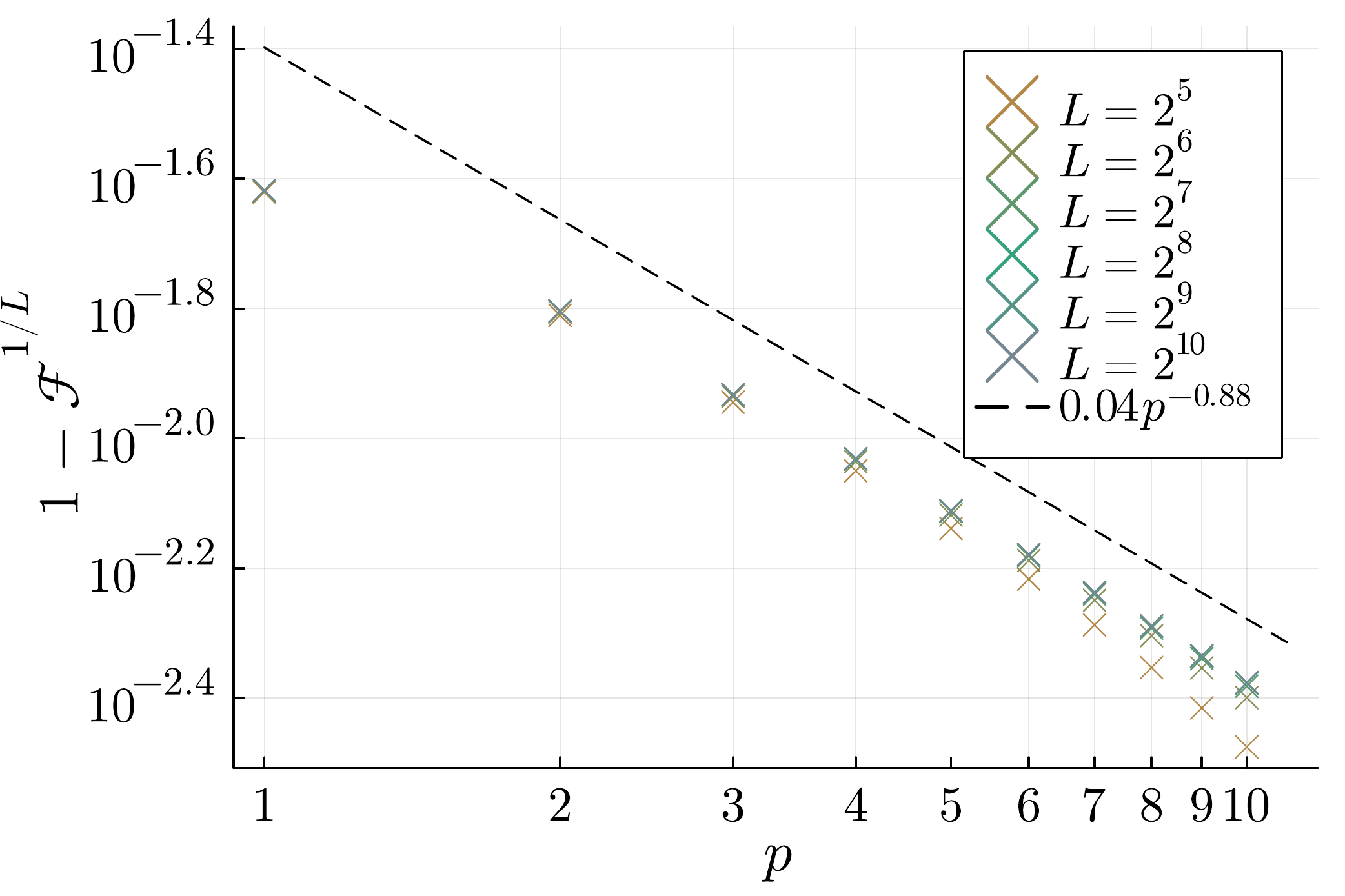}
  \caption{}
 \end{subfigure}
  \caption{\\ \textbf{(a):} Error of variational energy density from the infinite volume value of $-4/\pi$ against total circuit depth / number of rounds $p$ for systems $L > 2p$. \\ \textbf{(b):}  Normalized pure state infidelity versus number of rounds $p$, for various system sizes $L > 2p$.  Error in both the energy density and normalized infidelity for short depth ansatz states appear consistent with polynomial decay in $p$ for large system sizes. } 
\label{fig:qaoa-en-fid}
\end{figure}

The energy density and normalized ground state infidelity for a local QAOA circuit ansatz are shown in Fig. \ref{fig:qaoa-en-fid}. The circuits are initialized using parameters provided in \cite{Ho_2019} for preparing exact ground states for system sizes $L=2p$, and then the energy is optimized for systems with $L>2p$ so that the circuit does not see the whole system, and therefore cannot prepare the exact ground state. The resulting states give us an idea of how short depth local circuits can perform when preparing approximate ground states. Both errors in energy density and state fidelity appear consistent with polynomial decay in the circuit depth. This contrasts with the DMERA ansatz which achieves exponentially improving accuracy with increasing $D$.

\section{Discussion}
\label{s:discussion}

The critical free-fermion model is special in many ways, but we expect similar circuit architectures to be successful more broadly for states of local Hamiltonians. In these situations, locality and separation of length scales have been invaluable tools in physical analysis which can also be leveraged in designing good quantum circuits for preparing physically relevant states. Adapting to systems which are non-integrable could be done by allowing a more general set of local gates which goes beyond Gaussian fermion operations (i.e. universal rather than matchgate circuits). Additionally, states with finite correlation length $\xi$ could be prepared using $O(\log \xi )$ scaling layers optimized independently, which may be interpreted as a nontrivial renormalization group flow which disentangles the state at scales beyond the correlation length. The variational ansatz states we find are able to outperform similar analytic constructions while maintaining much of their simplicity, compared to a more general variational ansatz such as a large bond dimension MERA state. Finding a way to more precisely formalize variational multi-scale circuits as genuine scaling transformations beyond state preparation may also find use in algorithms for simulating conformal field theories and quantum field theories using closely ideas related from renormalization and wavelet theory \cite{CFTsim, Brennen_2015, Bagherimehrab_2022, Witteveen_2021, Lattice_ferm}. For NISQ applications, mitigation of error under tight hardware constraints is a key consideration. The symmetry averaging methods described here for mitigating ansatz-induced error complement prior work showing that DMERA's renormalization structure provides intrinsic resilience against hardware-induced error \cite{DMERA, ERprep}. \\
\\
\textbf{Acknowledgements:} The authors thank Brian Swingle and Christopher White for useful discussions. N.B. is funded by the Spin Chain Bootstrap for Quantum Computation project from the DOE Office of Science- Basic Energy Sciences, project number PM602. T.J.S. is grateful for support from the Department of Energy under award number DESC0019139.

\bibliographystyle{unsrt} 

\bibliography{refs.bib}

\appendix

\section{Free Fermions}

Some basics of the representation of Gaussian fermionic states and operations are outlined here following \cite{Bravyi_2017}. See also the Grassmann representation in \cite{FLO}. 

A Gaussian fermion state $\rho$ can be fully characterized by the covariance matrix
\begin{align}
  \label{eq:covmat}
  \begin{split}
    \Gamma_{j,k} = \frac{i}{2} \mathrm{tr}( \rho [\gamma_j, \gamma_k])
    \end{split}
\end{align}

For $n$ local fermionic modes $a_i$ we can define $2n$ Majorana modes 

\begin{align}
\label{eq:maj}
\gamma_{2i-1} &= \frac{a_i + a^\dagger_i}{2} \\
\gamma_{2i} &=  \frac{a_i - a^\dagger_i}{2} 
\end{align} 

These operators are Hermitian and obey the usual fermionic anticommutation relations ${\gamma_i,\gamma_j} = 2I$, but  also $\gamma^2_i = \frac{I}{2}$.

A Gaussian fermion state satisfies $\Gamma^2 \leq -I$, with equality holding for pure states. The expectation value of quadratic fermion observables can be directly read from the covariance matrix, while Wick's theorem allows higher order observables to be calculated in terms of the quadratic expectation values. 

Subsystems of a Gaussian state are defined by restricting the covariance matrix to a subset of the Majorana operators, eliminating the rows and columns associated with the discarded operators. Any mixed Gaussian state can be considered a subsystem of some pure Gaussian state. The covariance matrix for a mixed state can be purified by adding additional Majorana modes, introducing the new rows and columns with entries so that the new covariance matrix satisfies $\Gamma^2 = -I$.

Although we can relate the fermionic system to a system of qubits using a Jordan-Wigner duality, this operation of taking fermionic subsystems is a different restriction of the Hilbert space from the tracing out of qubit degrees of freedom. However, all even parity operators on the reduced fermionic and qubit subsystems will agree. Odd parity operators have zero expectation value in Gaussian states. 

A Gaussian unitary transformation always exists which transforms a Gaussian state into a collection of independent fermion modes, each with excitation probability $p_j$, so every Gaussian state has a product spectrum.
 Gaussian unitary operations are equivalent to linear transformations on the fermion modes, which act on the covariance matrix via an orthogonal transformation $\Gamma' = O \Gamma O^T$. 
The Gaussian unitary transformation that decouples the Gaussian fermion state block diagonalizes the covariance matrix into two-by-two blocks with off diagonal entries $\pm i \lambda_j$. 

Because all Gaussian states have a product spectrum, the entropy of any mixed Gaussian state is the sum of entropies $S_j$ from the states eigenmodes, each of which can be written in terms of the respective eigenvalue of the covariance matrix. 

 \begin{align}
  \label{eq:covEnt}
  \begin{split}
   S_j &=  -  \frac{1 + \lambda_j }{2} \log \frac{1 + \lambda_j }{2} -  \frac{1 - \lambda_j }{2} \log \frac{1 - \lambda_j }{2} \\
   &= \log 2 - \log \sqrt{1 - \lambda_j^2} - \lambda_j \log \sqrt{\frac{1 + \lambda_j}{1 - \lambda_j}}
    \end{split}
\end{align}

The fidelity between two mixed Gaussian states can also be computed in terms of their covariance matrices \cite{SW19}, 

 \begin{align}
  \label{eq:covFid}
  \begin{split}
   \mathcal{F}_{\rho, \sigma} =  2^{-\frac{n}{2}} \mathrm{det}(I - \Gamma_\rho \Gamma_\sigma)^{\frac{1}{4}} \mathrm{det}\left ( I - \sqrt{ I + \Gamma_{\rho,\sigma}^2} \right )^{\frac{1}{4}}
    \end{split}
\end{align}
with $\Gamma_{\rho,\sigma} = \frac{\Gamma_\rho + \Gamma_\sigma}{I - \Gamma_\rho \Gamma_\sigma}$. 

If at least one of these states is pure then the simpler expression can be used, $ \mathcal{F}_{\rho, \sigma} =  |\det( (\Gamma_\rho + \Gamma_\sigma)/2)|^{1/4}$.

The quadratic fermionic Hamiltonian we work with which is dual to the critical transverse-field Ising model is:

\begin{align}
  \label{eq:ham}
  \begin{split}
    H &=   i \sum_{j=1}^{2L} \gamma_j \gamma_{j+1}
  \end{split}
\end{align}

For a finite sized spin chain with periodic boundary conditions, the dual quadratic fermion Hamiltonian must have anti-periodic boundary conditions for the even parity states, and periodic boundary conditions for odd parity states. Thus, with periodic boundary conditions the full Ising Hamiltonian is not strictly dual to a single quadratic fermion Hamiltonian, but all of its eigenstates are dual to a Gaussian fermion states.

To avoid these issues we will focus just on the Majorana fermion model with anti-periodic boundary conditions for our numerics, and in this case our circuits may be interpreted as local operations on fermionic degrees of freedom. In fact, because we are working with a free-fermion model, we will also restrict our gates to be unitary Gaussian fermion operations, equivalent to Bogoliubov transformations.
 
Because local matchgate circuits are equivalent to linear transformations on the fermionic modes, this subgroup of quantum states and operations and can be simulated by classical polynomial time algorithms \cite{Jozsa_2008}, and may implemented numerically using covariance matrix methods or via fermionic linear optics \cite{FLO}. Under Jordan-Wigner duality, local two-qubit gates become orthogonal transformations on four consecutive Majorana operators. These techniques can be extended to tensor networks as matchgate tensor networks \cite{Schuch_2019,Jahn_2017}.
  
The two-qubit gates $u(x,y)$ can be written implemented as an orthogonal transformations $\tilde{u}(x', y')$ on four Majorana operators, using the parameters $x' = x+y$ and $y' = x-y$ as in eq. \eqref{eq:u-tilde}.

\begin{equation}
\tilde{u}(x',y') \equiv \begin{bmatrix}
\cos(x') & 0 & - \sin(x')& 0 \\
 0& \cos(y') & 0 & \sin(y')  \\
  \sin(x')& 0 & \cos(x') & 0 \\
 0& -\sin(y') & 0 & \cos(y') \\

\end{bmatrix}
\label{eq:u-tilde}
\end{equation}

The variational parameters found and used in this paper are reported as $\theta_D$ for a depth $D$ ansatz with $2D$ parameters, see eq. \eqref{eq:params}. Here, the $(2i -1)^{th}$ parameter is the $x'$ parameter for the $i^{th}$ layer of gates within the scaling transformation and the $(2i -1)^{th}$ parameter is the $y'$ parameter for the same layer of gates. 

We also found alternate sets of parameters for the modified Ising model in Eq. \eqref{eq:ham3}, listed in \eqref{eq:params2}. These circuits have similar features and the symmetry-averaging improvement is displayed in Fig. \ref{fig:sym-ratio} along with those of the original Ising model states. 

\begin{widetext}
\begin{align}
  \label{eq:params}
  \begin{split}
   \theta_1 &= [0.43188, \hspace{2mm} -1.13891 ] \\
 \theta_2 &=  [0.1379, \hspace{2mm} -0.56374, \hspace{2mm} -0.53456, \hspace{2mm} 0.18071]\\
  \theta_3 &= [-1.79716, \hspace{2mm} -1.51891, \hspace{2mm} 0.64486,\hspace{2mm} 2.0904, \hspace{2mm} 0.10994, \hspace{2mm} -0.31494]\\
 \theta_4 &=  [-1.66528, \hspace{2mm} -1.55101, \hspace{2mm} 1.05114, \hspace{2mm} 1.82904,\\ & \hspace{5mm}  0.43426, \hspace{2mm} -0.74951, \hspace{2mm} 0.07349, \hspace{2mm} -0.20764]\\
  \theta_5 &= [-1.61046, \hspace{2mm} -1.56358, \hspace{2mm} 1.29107, \hspace{2mm} 1.69499, \hspace{2mm} 0.80132,\\ & \hspace{5mm}  \hspace{2mm} -1.06696, \hspace{2mm} 0.30863, \hspace{2mm} -0.54539, \hspace{2mm} 0.05158, \hspace{2mm} -0.14651]\\
  \theta_6 &= [-1.58682, \hspace{2mm} -1.56831, \hspace{2mm} 1.42666, \hspace{2mm} 1.6278,\hspace{2mm}  1.08403,\hspace{2mm}  -1.28163, \\ & \hspace{5mm} \hspace{2mm}  0.62053,\hspace{2mm}  -0.85436,\hspace{2mm}  0.2323, \hspace{2mm} -0.4140,\hspace{2mm}  0.03862,\hspace{2mm}  -0.11017]\\
  \end{split}
\end{align}

\begin{align}
  \label{eq:params2}
  \begin{split}
    \tilde{\theta}_1 &= [-0.22107 \hspace{2mm} -1.79187 ] \\
 \tilde{\theta}_2 &=  [0.37921, \hspace{2mm} -1.58672, \hspace{2mm} -0.23588, \hspace{2mm}  -0.8906]\\
  \tilde{\theta}_3 &= [0.09744, \hspace{2mm} -1.62352, \hspace{2mm}  -0.86247,\hspace{2mm} -0.29472, \hspace{2mm}  -0.18693, \hspace{2mm}  0.51259]\\
  \tilde{\theta}_4 &=  [-0.0058, \hspace{2mm} -0.21063, \hspace{2mm} 1.6268, \hspace{2mm} 0.78319, \\ & \hspace{5mm} -1.08715,\hspace{2mm}  0.1295,\hspace{2mm}  2.77886, \hspace{2mm} -0.08755]\\
   \tilde{\theta}_5 &= [-0.08357, \hspace{2mm} -1.73696, \hspace{2mm}  -0.09269,\hspace{2mm}  -0.14014, \hspace{2mm} 0.14162, \\ & \hspace{5mm} 1.32826,\hspace{2mm}  -1.05885,\hspace{2mm}  0.11338,\hspace{2mm}  -0.42156,\hspace{2mm}  -0.40656]\\
   \tilde{\theta}_6 &= [0.03614, \hspace{2mm} -1.56255, \hspace{2mm} -1.11474, \hspace{2mm} -0.28192, \hspace{2mm} 0.55476, \hspace{2mm} 2.59916, \\ & \hspace{5mm} -0.08826,\hspace{2mm}  0.10361,\hspace{2mm}  -1.05898, \hspace{2mm} -1.74665,\hspace{2mm}  -0.05094,\hspace{2mm}  0.25421]\\
  \end{split}
\end{align}
\end{widetext}

\end{document}